\documentclass[]{INCLUDES/llncs}
\usepackage{amsmath}
\usepackage{graphicx}
\usepackage{subfig}
\usepackage{url}
\usepackage{verbatim} 
\usepackage{listings}
\lstnewenvironment{code}
    {\lstset{}%
      \csname lst@SetFirstLabel\endcsname}
    {\csname lst@SaveFirstLabel\endcsname}
\def \bscale1 {1.00}
\def \bscale {0.20}

\begin{document}

\title{
   Logic Engines as Interactors
}

\author{Paul Tarau}
\institute{
   Department of Computer Science and Engineering\\
   University of North Texas\\
   {\em E-mail: tarau@cs.unt.edu}
}

\maketitle

\date{}

\begin{abstract}
We introduce a new programming language construct, {\em Interactors}, 
supporting the agent-oriented view that programming is a dialog between 
simple, self-contained, autonomous building blocks. 

We define {\em Interactors} as an abstraction of answer generation 
and refinement in {\em Logic Engines} resulting in expressive 
language extension and metaprogramming patterns, including 
emulation of Prolog's dynamic database. 

A mapping between backtracking based answer generation in 
the callee and ``forward" recursion in the caller enables
interaction between different branches of the callee's search process 
and provides simplified design patterns for
algorithms involving combinatorial generation and 
infinite answer streams.

Interactors extend language constructs like Ruby, Python 
and \verb~C#~'s multiple coroutining block returns through 
{\em yield} statements and they can emulate the action 
of monadic constructs and catamorphisms in functional languages.

{\em {\bf Keywords}:
generalized iterators, 
logic engines, 
agent oriented programming language constructs,
interoperation with stateful objects,
metaprogramming}
\end{abstract}

\section{Introduction}

Interruptible Iterators are 
a new Java extension described in 
\cite{DBLP:conf/popl/LiuKM06,DBLP:conf/padl/LiuM03}. 
The underlying construct is the {\tt yield} statement 
providing multiple returns and resumption of iterative 
blocks. It has been integrated in newer Object Oriented 
languages  like 
{\tt Ruby} \cite{matz,DBLP:conf/oopsla/Sasada05} 
\verb~C#~ \cite{CSharp} and {\tt Python} \cite{DBLP:conf/tools/Rossum97} 
but it goes back to the {\em Coroutine Iterators} introduced in older 
languages like CLU \cite{DBLP:books/sp/Liskov81} and 
ICON \cite{DBLP:journals/toplas/GriswoldHK81}.

Our next stepping stone is the more radical idea of allowing clients to 
communicate to/from inside blocks of arbitrary recursive computations. 
The challenge is to achieve this without the fairly complex interrupt based
communication protocol between the iterator and its client 
described in \cite{DBLP:conf/popl/LiuKM06}.
As a natural generalization, the need arises for a 
structured two-way communication between a client and the usually 
autonomous service the client requires from a 
given language construct, often encapsulating
an independent component.

Agent programming constructs have influenced design patterns 
at ``macro level'', ranging from interactive Web services to 
mixed initiative computer human interaction. {\em Performatives} 
in Agent communication languages 
\cite{DBLP:conf/atal/MayfieldLF95,fipa2:97} 
have made these constructs reflect explicitly the intentionality, 
as well as the negotiation process involved in agent interactions. 
At a more theoretical level, it has been argued 
that {\em interactivity}, seen as fundamental computational paradigm, 
can actually expand computational expressiveness and provide 
new models of computation \cite{DBLP:journals/cj/WegnerE04}.

In a logic programming context, the Jinni agent programming 
language  \cite{ciclops:jinni,j2k_ug} and the BinProlog 
system \cite{bp7advanced} have been centered 
around logic engine constructs providing an API that supported 
reentrant instances of the language processor. 
This has naturally led to a view of logic engines as instances 
of a generalized family of iterators called {\em Fluents} 
\cite{tarau:cl2000}, that have allowed the separation of the 
first-order language interpreters from the multi-threading 
mechanism, while providing a very concise source-level 
reconstruction of Prolog's built-ins.

Building upon the {\em Fluents} API described in \cite{tarau:cl2000}, 
this paper will focus on bringing interaction-centered, 
agent oriented constructs from software design frameworks 
and design patterns to programming language level.

The resulting language constructs, that we shall call {\em Interactors}, 
will express control, metaprogramming and interoperation with 
stateful objects and external services.
They complement pure Horn Clause Prolog with a significant
boost in expressiveness, to the point where they
allow emulating at source level virtually all Prolog
builtins, including dynamic database operations.

As paradigm independent language constructs, {\em Interactors} 
are a generalization of 
{\em Coroutine Iterators} \cite{DBLP:books/sp/Liskov81}
and 
{\em Interruptible Iterators} \cite{DBLP:conf/popl/LiuKM06}.

Their extra expressiveness comes from the fact that they 
embed arbitrarily nested computations and provide linearized 
data exchange mechanisms with their internal components. 
In particular, a mapping between backtracking based answer 
generation in the callee, and forward recursion in the 
caller, facilitating interaction between different branches 
of the callee's search process, becomes possible. 

As the operation of interrupting iterators 
already has a fairly complex 
semantics, interrupting general computational process is 
even trickier and more likely to be unsafe, as their state is 
unknown to the clients. Therefore, Interactors are designed 
to support cooperative data exchanges rather then 
arbitrary interrupts. 

Independently, the need for {\em state representation 
with minimal new ontology}
in declarative languages arises from seeking
simplified {\em interoperation} with I/O and conventional 
software and operating system services that often 
relay on stateful entities. 
In this sense, Interactors solve for Logic Programming 
languages problems similar to the problems that
constructs like Monads and Catamorphisms 
solve for Functional Languages 
\cite{DBLP:conf/mpc/BirdM92,DBLP:journals/jfp/Hutton99}.

\section{First Class Logic Engines} \label{logeng}

To make the paper self-contained, we will start with an 
overview of the {\em Logic Engine API} 
introduced in \cite{tarau:cl2000} to which 
our {\em Interactor API} is a natural extension.

\subsection{Engines as a Reflection Layer}

Speaking generically, an {\em Engine} is simply a 
language processor reflected through an API that 
allows its computations to be controlled 
interactively from another {\em Engine}
very much the same way a programmer controls Prolog's interactive 
toplevel loop: launch a new goal, ask for a new answer, 
interpret it, react to it. 

A {\em Logic Engine} is an {\em Engine} 
running a Horn Clause Interpreter 
with LD-resolution \cite{Tarau93:CONS,Tarau91:JAP} 
on a given clause database, together with a set of 
built-in operations.

Each {\em Logic Engine} has a constructor which initializes it 
with a {\em goal} and an {\em answer pattern}.
In fact, an engine can be seen as a 
generator of a (possibly infinite) 
{\em stream of answers} which 
can be explored one by one, i.e. an {\em Iterator} 
over a stream of answers.
To use a simple analogy, the object encapsulating 
the state of the runtime interpreter is very similar to
a file descriptor encapsulating the advancement 
of a file reader.

Logic Engines will have the ability to create and query 
other Logic Engines, as part of a general mechanism 
to manipulate {\tt Interactors}.

{\tt Interactors} encapsulating logic engines, like any other 
stateful objects, will have their independent life-cycles.
This general mechanism will allow Logic Engines 
to interoperate with the underlying imperative implementation language,
which provides them and requests from them
various services through a hierarchy of {\tt Interactors}.

Each Logic Engine based Interactor works as a separate 
Horn Clause LD-resolution interpreter. The engine constructor, 
when called, initializes a new, lightweight interpreter, having
its own stacks and heap.

The command
\begin{verbatim}
new_engine(AnswerPattern,Goal,Interactor)
\end{verbatim}

{\noindent  creates} a new Horn Clause solver, uniquely identified
by {\tt Interactor}, which shares code
with the currently running program and is initialized
with {\tt Goal} as a starting point. 
{\tt AnswerPattern} is a term, usually
a list of variables occurring in {\tt Goal},
of which answers
returned by the engine will be instances.

The {\tt get/2} operation is used to retrieve successive 
answers generated by an Interactor, on demand.

\begin{verbatim}
get(Interactor,AnswerInstance)
\end{verbatim}

\noindent  It tries to harvest the answer computed from {\tt Goal}, 
as an instance of {\tt AnswerPattern}. If an answer
is found, it is returned as {\tt the(AnswerInstance)}, 
otherwise the atom {\tt no} 
is returned. Note that once the atom {\tt no} has been
returned, all subsequent {\tt get/2} operations on the 
same {\tt Interactor} will return {\tt no}.
As in the case of {\tt Maybe} Monad in Haskell, 
returning distinct functors in the case of success and 
failure, allows
further case analysis in a pure Horn Clause style, 
without needing Prolog's CUT or if-then-else operation. 

Note that bindings are not propagated to the original {\tt Goal}
or {\tt AnswerPattern} when {\tt get/2} retrieves an answer,
i.e. {\tt AnswerInstance} is obtained by first standardizing apart
(renaming) the variables in {\tt Goal} and {\tt AnswerPattern}, and then
backtracking over its alternative answers in a separate Prolog
interpreter. Therefore, backtracking in the caller interpreter does
not interfere with the new Interactor's iteration over answers. 
Backtracking over the Interactor's creation point, as such, 
makes it unreachable and therefore subject to garbage collection.

An Interactor is stopped with the {\tt stop/1} operation 
(that is also called automatically when no more answers 
can be produced):

\begin{verbatim}
stop(Interactor)
\end{verbatim}

\noindent So far, these operations provide a minimal {\em Coroutine 
Iterator API}, powerful
enough to switch tasks cooperatively between an engine and its client
and emulate key Prolog built-ins like
{\tt if-then-else} and {\tt findall} \cite{tarau:cl2000},
as well as other higher order operations 
similar to Haskell's {\em fold}  (subsection \ref{higher}).

These interactor operations correspond  to the {\em Answer Source} 
fluents described in \cite{tarau:cl2000}, where a complete 
specification of their operational semantics is given.

\section{From Fluents to Interactors} \label{interactors}

We will now describe the extension of the {\em Fluents} 
API of \cite{tarau:cl2000} that provides a minimal bidirectional 
communication API between interactors and their clients.

\subsection{The Interaction Mechanism} \label{intmec}

The following operations provide a ``mixed-initiative'' 
interaction mechanism, allowing more general data exchanges 
between an engine and its client.

\subsubsection{A yield/return operation} \label{return}

First, like the {\tt yield return} construct of \verb~C#~ 
and the {\tt yield operation} of Ruby and Python, 
our {\tt return/1} operation 

\begin{verbatim}
return(Term)
\end{verbatim}

\noindent will save the state of the engine and transfer 
control and a result {\tt Term} to its client. The client 
will receive a copy of {\tt Term} simply by using 
its {\tt get/1} operation. 
Similarly to Ruby's {\tt yield}, our {\tt return} 
operation suspends and returns data from arbitrary 
computations (possibly involving recursion) rather than 
from specific language constructs like a {\tt while} 
or {\tt for} loop.

Note that an Interactor returns control to its client 
either by calling {\tt return/1}
or when a computed answer becomes available.
By using a sequence of {\tt return/get} operations, 
an engine can provide a stream of {\em intermediate/final results} 
to its client, without having to backtrack. This mechanism 
is powerful enough to implement a complete exception 
handling mechanism (see \cite{tarau:cl2000}) simply with

\begin{verbatim}
throw(E):-return(exception(E)).
\end{verbatim}

\noindent When combined with a {\tt catch(Goal,Exception,OnException)}, 
on the client side, the client can decide,
upon reading the exception with {\tt get/1}, if it wants 
to handle it or to throw it to the next level.

The mechanisms discussed so far are expressive enough, 
as described in \cite{tarau:cl2000}, to implement at source 
level key built-in predicates of Prolog like 
{\tt if-then-else, findall} and {\tt copy\_term}.

\subsubsection{Interactors and Coroutining} \label{yield}
  
The operations described so far allow an engine to return 
answers from any point in its computation sequence. 
The next step is to enable its client to {\em inject} 
new goals (executable data) to an arbitrary inner context 
of an engine. Two new primitives are needed:

\begin{verbatim}
to_engine(Engine,Data)
\end{verbatim}
\noindent used to send a client's data to an Engine, and

\begin{verbatim}
from_engine(Data)
\end{verbatim}
used by the engine to receive a client's Data.

Using a metacall mechanism like {\tt call/1} (which can also be emulated
in terms of engine operations \cite{tarau:cl2000}), one can implement
a close equivalent of Ruby's {\tt yield} statement as follows:

\begin{verbatim}
ask_engine(Engine,Goal, Answer):-
  to_engine(Engine,Goal),
  get(Engine,Answer).

engine_yield(Answer):-
  from_engine((Answer:-Goal)),
  call(Goal),
  return(Answer).
\end{verbatim}

\noindent where {\tt ask\_engine} sends a goal (possibly built 
at runtime) to an engine, which in turn, executes it and returns 
a result with an {\tt engine\_yield} operation. 

As the following example shows, this allows the client 
to use from outside the (infinite) recursive loop of an engine 
as a form of {\em updatable persistent state}.

\begin{verbatim}
sum_loop(S1):-engine_yield(S1=>S2),sum_loop(S2).

inc_test(R1,R2):-
   new_engine(_,sum_loop(0),E),
   ask_engine(E,(S1=>S2:-S2 is S1+2),R1),
   ask_engine(E,(S1=>S2:-S2 is S1+5),R2).
   
?- inc_test(R1,R2).
R1=the(0 => 2),
R2=the(2 => 7)
\end{verbatim}

\noindent  Note also that after parameters (the increments 2 and 5)
are passed to the engine, results dependent on its state 
(the sums so far 2 and 7) are received back. Moreover, note 
that an arbitrary goal is injected in the local context of the 
engine where it is executed, with access to the engine's 
{\em state variables} {\tt S1} and {\tt S2}. 
As engines have separate garbage collectors (or in simple 
cases as a result of tail recursion), their infinite loops 
run in constant space, provided that no unbounded size objects 
are created.

We will call {\tt Interactors API} the Horn Clause subset of Prolog 
with LD resolution together with the Logic Engine operations 
described so far. As we shown in \cite{tarau:cl2000}, {\tt call/1} 
itself can be emulated at source level with the {\tt Logic Engine API}.
As shown in subsection \ref{db}, the API will also allow emulating 
Prolog's dynamic database operations, providing runtime code creation 
and execution.

\subsection{Using Interactors}

To summarize, a typical use case for the {\em Interactor API} looks as follows:

\begin{enumerate}
\item the {\em client} creates and initializes a new {\em engine}
\item the client triggers a new computation in the {\em engine}, 
parameterized as follows:

\begin{enumerate}
\item the {\em client} passes some data and a new goal to the {\em engine} 
and issues a {\tt get} operation that passes control to it
\item the {\em engine} starts a computation from its initial goal or 
the point where it has been suspended and runs (a copy of) 
the new goal received from its {\em client}
\item the {\em engine} returns (a copy of) the answer, 
then suspends and returns control to its {\em client}
\end{enumerate}
\item the {\em client} interprets the answer and proceeds 
with its next computation step
\item the process is fully reentrant and the {\em client} 
may repeat it from an arbitrary point in its computation
\item while cooperation between the {\em engine} and its 
client is assumed, the ``{\em client} drives'' ; failure 
of the goal injected in the {\em engine}'s computation space 
fails the {\em engine} (``fast fail'' semantics)
\end{enumerate}

A number of alternate semantics are possible, implementable at 
source level, on top of {\tt to\_engine/2} and {\tt from\_engine/1}. 
An alternate scenario, emphasizing more on error recovery than 
``fast fail'' could be devised: when failure and/or exceptions are 
caught by the engine, the client is simply notified about them, 
while the engine's ability to handle future requests is preserved.

\subsection{Interactors and Higher Order Constructs} \label{higher}
As a first glimpse at the expressiveness of this API, we will implement, 
in the tradition of higher order functional programming, 
a {\em fold} operation 
\cite{DBLP:conf/mpc/BirdM92} 
connecting results produced by independent branches 
of a backtracking Prolog engine:

\begin{verbatim}
efoldl(Engine,F,R1,R2):-
  get(Engine,X),
  efoldl_cont(X,Engine,F,R1,R2).

efoldl_cont(no,_Engine,_F,R,R).
efoldl_cont(the(X),Engine,F,R1,R2):-
  call(F,R1,X,R),
  efoldl(Engine,F,R,R2).
\end{verbatim}

\noindent Classic functional programming idioms like 
{\em reverse as fold} are then implemented simply as: 

\begin{verbatim} 
reverse(Xs,Ys):-
  new_engine(X,member(X,Xs),E),
  efoldl(E,reverse_cons,[],Ys).  
  
reverse_cons(Y,X,[X|Y]).
\end{verbatim}

Note also the automatic {\em deforestation} effect 
\cite{journals/tcs/Wadler90,conf/fp/MarlowW92} of this programming 
style - no intermediate list structures need to be built, 
if one wants to aggregate the values retrieved from 
an arbitrary generator engine with an operation 
like sum or product.

\section{Interactors and Interoperation with Stateful Objects}

The gain in expressiveness coming directly from the view of 
logic engines as answer generators is significant. 
We refer to \cite{tarau:cl2000} for source level implementations of
virtually all essential Prolog built-ins (exceptions included). 
The notable exception is Prolog's dynamic database, 
requiring the bidirectional communication provided by interactors.

\subsection{Dynamic Databases with Interactors} \label{db}

The key idea for implementing dynamic database operations 
with Interactors is to use a logic engine's state in an 
infinite recursive loop, similar to the coinductive 
programming style advocated in 
\cite{DBLP:conf/icalp/SimonBMG07,DBLP:conf/iclp/GuptaBMSM07}, 
to emulate state changes in its client engine.

First, a simple difference-list based infinite server loop is built:

\begin{verbatim}
queue_server:-queue_server(Xs,Xs).
    
queue_server(Hs1,Ts1):-
  from_engine(Q),
    server_task(Q,Hs1,Ts1,Hs2,Ts2,A),
  return(A),
  queue_server(Hs2,Ts2).
\end{verbatim}

\noindent Next we provide the queue operations, 
needed to maintain the state of the database.

\begin{verbatim}
server_task(add_element(X),Xs,[X|Ys],Xs,Ys,yes).
server_task(push_element(X),Xs,Ys,[X|Xs],Ys,yes).
server_task(queue,Xs,Ys,Xs,Ys,Xs-Ys).
server_task(delete_element(X),Xs,Ys,NewXs,Ys,YesNo):-
  server_task_delete(X,Xs,NewXs,YesNo).
\end{verbatim}

\noindent Then we implement the auxiliary predicates 
supporting various queue operations:
\begin{verbatim}
server_task_remove(Xs,NewXs,YesNo):-nonvar(Xs),Xs=[X|NewXs],!,
  YesNo=yes(X).
server_task_remove(Xs,Xs,no).

server_task_delete(X,Xs,NewXs,YesNo):-select_nonvar(X,Xs,NewXs),!,
  YesNo=yes(X).
server_task_delete(_,Xs,Xs,no).

server_task_stop(E):-stop(E).

select_nonvar(X,XXs,Xs):-nonvar(XXs),XXs=[X|Xs].
select_nonvar(X,YXs,[Y|Ys]):-nonvar(YXs),YXs=[Y|Xs],
  select_nonvar(X,Xs,Ys).
\end{verbatim}

\noindent Finally, we put it all together, 
as a dynamic database API:

\begin{verbatim}
% creates a new engine server 
% providing Prolog database operations
new_edb(Engine):-new_engine(done,queue_server,Engine).

% adds an element to the end of the database
edb_assertz(Engine,Clause):-
  ask_engine(Engine,add_element(Clause),the(yes)).
  
% adds an element to the front
edb_asserta(Engine,Clause):-
  ask_engine(Engine,push_element(Clause),the(yes)).
  
% returns a instances of asserted clauses
edb_clause(Engine,Head,Body):-
  ask_engine(Engine,queue,the(Xs-[])),
  member((Head:-Body),Xs).
  
% delete an element of the database
edb_retract1(Engine,Head):-Clause=(Head:-_Body),
  ask_engine(Engine,delete_element(Clause),the(yes(Clause))).
  
% removes a database
edb_delete(Engine):-stop(Engine).
\end{verbatim}
\noindent The database will now generate the equivalent 
of {\tt clause/2}, ready to be passed to a Prolog metainterpreter.
\begin{verbatim}
test_clause(Head,Body):-
  new_edb(Db),
    edb_assertz(Db,(a(2):-true)),
    edb_asserta(Db,(a(1):-true)),
    edb_assertz(Db,(b(X):-a(X))),
  edb_clause(Db,Head,Body).
\end{verbatim}
  
\noindent Externally implemented dynamic databases 
are also made visible as {\tt Interactors} and reflection 
of the interpreter's own handling of the Prolog 
database becomes possible. 
As an additional benefit, multiple databases are provided.
This simplifies adding module, object or agent layers at source level. 
By combining database and communication 
(socket or RMI) Interactors, software abstractions
like mobile code and autonomous agents are built 
as shown in \cite{td:tlp}.
Interoperation with External Stateful Objects like file 
systems or Prolog language extensions as dynamic databases 
is also simpler as implementation language operations 
can be applied to Interactors directly.
Moreover, Prolog operations traditionally
captive to predefined list based implementations (like DCGs)
can be made generic and mapped to work directly 
on Interactors encapsulating file, URL and socket Readers.

\section{Refining Control and Simplifying Algorithms with Interactors}

\subsection{Refining control: a backtracking if-then-else}
Modern Prolog implementations (SWI, SICStus, BinProlog) also provide
a variant of {\tt if-then-else} that either backtracks 
over multiple answers of its
{\tt then} branch or switches to the {\tt else} branch 
if no answers in the {\tt then} branch are found. 
With the same API, we can implement it at source level as follows
\footnote{We have included this example because it expresses a 
form of control that cannot be implemented at source level. 
Although discussed in a posting of the author in comp.lang.prolog,
this example has never been part of a reviewed publication.}:

\begin{verbatim}
if_any(Cond,Then,Else):-
  new_engine(Cond,Cond,Engine),
  get(Engine,Answer),
  select_then_or_else(Answer,Engine,Cond,Then,Else).

select_then_or_else(no,_,_,_,Else):-Else.
select_then_or_else(the(BoundCond),Engine,Cond,Then,_):-
  backtrack_over_then(BoundCond,Engine,Cond,Then).

backtrack_over_then(Cond,_,Cond,Then):-Then.
backtrack_over_then(_,Engine,Cond,Then):-
  get(Engine,the(NewBoundCond)),
  backtrack_over_then(NewBoundCond,Engine,Cond,Then).
\end{verbatim}

\subsection{Simplifying Algorithms: 
Interactors and Combinatorial Generation}

Various combinatorial generation algorithms have elegant 
backtracking implementations. However, it is notoriously 
difficult (or inelegant, through the use of impure side effects) 
to compare answers generated by different OR-branches 
of Prolog's search tree.

\subsubsection{Comparing Alternative Answers}
Such optimization problems can easily be expressed as follows:
\begin{itemize}
\item running the generator in a separate logic engine
\item collecting and comparing the answers in a client controlling the engine
\end{itemize}
\noindent The second step can actually be automated, provided that 
the comparison criterion is given as a predicate 
\begin{verbatim}
compare_answers(First,Second,Best)
\end{verbatim}
\noindent to be applied to the engine with an {\tt efold} operation
\begin{verbatim}
best_of(Answer,Comparator,Generator):-
  new_engine(Answer,Generator,E),
  efoldl(E,
    compare_answers(Comparator),no,
  Best),
  Answer=Best.

compare_answers(Comparator,A1,A2,Best):-
  if((A1\==no,call(Comparator,A1,A2)),
    Best=A1,
    Best=A2
  ).

?-best_of(X,>,member(X,[2,1,4,3])).
X=4
\end{verbatim}

\subsubsection{Counting Answers without Accumulating}
Problems as simple as counting the number of solutions of a 
combinatorial generation problem can become tricky in 
Prolog (unless one uses impure side effects) as one might 
run out of space by having to generate all solutions 
as a list, just to be able to count them. 
The following example shows how this can be achieved using 
an {\tt efold} operation on an integer partition generator:
\begin{verbatim}
integer_partition_of(N,Ps):-
  positive_ints(N,Is),
  split_to_sum(N,Is,Ps).

split_to_sum(0,_,[]).
split_to_sum(N,[K|Ks],R):-N>0,sum_choice(N,K,Ks,R).

sum_choice(N,K,Ks,[K|R]):-NK is N-K,split_to_sum(NK,[K|Ks],R).
sum_choice(N,_,Ks,R):-split_to_sum(N,Ks,R).

positive_ints(1,[1]).
positive_ints(N,[N|Ns]):-N>1,N1 is N-1,positive_ints(N1,Ns).

% counts partitions by running 
% the generator on an engine that returns 
% 1 for each answer that is found
count_partitions(N,R):-
  new_engine(1,
    integer_partition_of(N,_),Engine),
  efoldl(Engine,+,0,R).
\end{verbatim}

\subsection{Encapsulating Infinite Computations Streams}
An infinite stream of natural numbers is implemented 
simply as:
\begin{verbatim}
loop(N):-return(N),N1 is N+1,loop(N1).
\end{verbatim}

The following example shows a simple space efficient 
generator for the infinite stream of prime numbers:
\begin{verbatim}
prime(P):-prime_engine(E),element_of(E,P).

prime_engine(E):-new_engine(_,new_prime(1),E).

new_prime(N):-N1 is N+1,
  if(test_prime(N1),true,return(N1)),
  new_prime(N1).

test_prime(N):-M is integer(sqrt(N)),between(2,M,D),N mod D =:=0
\end{verbatim}
\noindent Note that the program has been wrapped, using 
the {\tt element\_of} predicate defined in \cite{tarau:cl2000}, 
to provide one answer at a time through backtracking. 
Alternatively, a forward recursing client can use 
the {\tt get(Engine)} operation to extract 
primes one at a time from the stream.

\section{Interactors and Multi-Threading}

While one can build a self-contained lightweight 
multi-threading API solely by switching control among 
a number of cooperating engines, with the advent of 
multi-core CPUs as the norm rather than 
the exception, the need for {\em native} multi-threading constructs 
is justified on both performance and expressiveness grounds. 
Assuming a dynamic implementation of a logic engine's stacks, 
Interactors provide lightweight independent 
computation states that can be easily mapped 
to the underlying native threading API. 

A minimal native Interactor based multi-threading API, 
has been implemented in the Jinni Prolog system \cite{j2k_ug} 
on top of a new thread launching built-in 
\begin{verbatim}
  run_bg(Engine,ThreadHandle)
\end{verbatim}.
\noindent This runs a new Thread starting from the engine's 
{\tt run()} method and returns a handle 
to the Thread object. To ensure that access to the 
Engine's state is safe and synchronized, we hide the engine 
handle and provide a simple producer/consumer data exchanger 
object, called a {\tt Hub}. The complete multi-threading 
API, partly designed to match Java's own threading API is:
\begin{itemize}
    \item {\tt bg(Goal)}: launches a new Prolog thread on 
    its own engine starting with {\tt Goal}.
    \item {\tt hub\_ms(Timeout,HubHandle)}: constructs a new 
    {\tt Hub} returned as a {\tt HubHandle} - a synchronization
    device on which N consumer threads can wait with {\tt collect(HubHandle,Data)} for data produced 
    by M producers providing data with 
    {\tt put(HubHandle,Data)}. However, if a given consumer 
    waits more than {\tt Timeout} milliseconds it returns and 
    signals failure. As usual in Java, {\tt Timeout=0} means 
    indefinite suspension. If the thread is meant to interact 
    with the parent, a {\tt HubHandle} can be given to it 
    as an argument.
    \item {\tt current\_thread(ThreadHandle)}: returns a handle 
    to the current thread - that might be passed to another 
    thread wanting to join this one.
    \item {\tt join\_thread(ThreadHandle)}: waits until 
    a given thread terminates.
    \item {sleep\_ms(Timeout)}: suspends current thread 
    for {\tt Timeout} milliseconds.
\end{itemize}

A number of advanced multi-threading libraries have
been designed around this basic API. For instance, 
{\em AND-synchronization} ensures waiting until N-tasks are finished. 
{\em Barriers} ensure that a number of threads wait jointly 
and when they all finish, a Runnable action is executed.

\paragraph{Associative Interactors}
The message passing style interaction shown in the previous 
sections between engines and their clients, can be easily 
generalized to associative communication through a unification 
based blackboard interface \cite{linda89,dbt95a}. 
Exploring this concept in depth promises more flexible 
interaction patterns, as out of order {\tt ask\_engine} 
and {\tt engine\_yield} operations would become possible, 
matched by association patterns.

\section{Interactors Beyond Logic Programming Languages}

We will now compare Interactors with similar constructs 
in other programming paradigms. 

\subsection{Interactors in Object Oriented Languages}

Extending Interactors to mainstream Object Oriented languages
is definitely of practical importance, given the gain in
expressiveness. For instance, in the implementation of
lightweight Prolog engines, Interactor based 
interfaces can provide a uniform interoperation mechanism
with most of the system built at source level.
Such an interface has been used in the Java-based Jinni Prolog 
system \cite{j2k_ug} to provide a uniform view of various 
Java services to Prolog's logic engines. In simple cases, 
like file operations, that can be suspended and resumed 
at will, such an interface is usable directly. 
In more complex cases, coroutining behavior can be achieved 
by adding {\tt switch/case} statements to keep track 
of advancement of the control flow in a 
method\footnote{Or just wait until Java borrows from 
{\tt Ruby} or {\tt C-sharp} something similar to the 
{\tt yield} or {\tt yield return} statements.}.
An elegant open
source Prolog engine {\tt Yield Prolog} has been
recently implemented in terms of Python's {\em yield} 
and \verb~C#~'s {\em yield return} primitives \cite{yieldProlog}. 
Extending Yield Prolog to support our Interactor API only 
requires adding the communication operations {\tt from\_engine} 
and {\tt to\_engine}. In older languages like 
{\tt Java}, \verb~C++~ or the {\tt Objective C} dialect 
used in Apple's iPhone SDK, one needs to implement a 
more complex API, 
including a {\tt yield return} emulation. 

\subsection{Interactors and similar constructs in 
Functional Languages}

Interactors based on logic engines encapsulate future 
computations that can be unrolled on demand. 
This is similar to lazy evaluation mechanisms 
in languages like Haskell \cite{haskell:98rr}.
Interactors share with 
Monads \cite{moggi:monads,wadler92:acm} 
the ability to sequentialize functional 
computations and encapsulate state information. 
As a minor detail, the returned values consisting 
of terms of the form {\tt the(Answer)} and {\tt no}, 
like the {\tt Maybe Monad}'s {\tt Just a} 
and {\tt Nothing} types, are used to encode 
possible failure of a computation. With higher 
order functions, monadic computations can pass 
functions to inner blocks. On the other hand, 
our {\tt ask\_engine} / {\tt engine\_yield} mechanism, 
like Ruby's {\tt yield}, is arguably more flexible, 
as it provides arbitrary switching of 
control (coroutining) between an Interactor and its client. 
Our ability to define Prolog's {\tt findall}  
construct \cite{tarau:cl2000} as well as {\tt fold} 
operations in terms of Interactors, is similar to 
definition of comprehensions 
\cite{wadler:comprehending:lfp:90,BT95a:ILPS} in 
terms of Monads, and List comprehensions in particular.

\section{Conclusion}

Logic Engines encapsulated as 
Interactors have been used to build on top of 
pure Prolog (together with the Fluent API described 
in \cite{tarau:cl2000}) a practical Prolog system, 
including dynamic database operations, entirely at 
source level. Interactors allowed 
to communicate between distinct OR-branches as a 
practical alternative to the use of
side effects and have provided elegant implementations 
of control structures and higher order predicates.

In a broader sense, Interactors can be seen as a starting 
point for rethinking fundamental programming language 
constructs like Iterators and Coroutining in terms 
of language constructs inspired by {\em performatives} 
in agent oriented programming. If the concept 
catches on, we expect it to impact on programmer 
productivity and simplification of software 
development at large. 

Beyond applications to 
logic-based language design, we hope that our 
language constructs will be reusable in the 
design and implementation of new functional 
and object oriented languages.

\bibliographystyle{plain}

\begin{thebibliography}{10}

\bibitem{BT95a:ILPS}
Yves Bekkers and Paul Tarau.
\newblock Monadic {C}onstructs for {L}ogic {P}rogramming.
\newblock In John Lloyd, editor, {\em Proceedings of ILPS'95}, pages 51--65,
  Portland, Oregon, December 1995. MIT Press.

\bibitem{DBLP:conf/mpc/BirdM92}
Richard~S. Bird and Oege de~Moor.
\newblock Solving optimisation problems with catamorphism.
\newblock In Richard~S. Bird, Carroll Morgan, and Jim Woodcock, editors, {\em
  MPC}, volume 669 of {\em Lecture Notes in Computer Science}, pages 45--66.
  Springer, 1992.

\bibitem{linda89}
N.~Carriero and D.~Gelernter.
\newblock {Linda in Context}.
\newblock {\em {CACM}}, 32(4):444--458, 1989.

\bibitem{dbt95a}
K.~De~Bosschere and P.~Tarau.
\newblock Blackboard-based {E}xtensions in {P}rolog.
\newblock {\em Software --- Practice and Experience}, 26(1):49--69, January
  1996.

\bibitem{fipa2:97}
FIPA.
\newblock {FIPA} 97 specification part 2: Agent communication language, October
  1997.
\newblock Version 2.0.

\bibitem{DBLP:journals/toplas/GriswoldHK81}
Ralph~E. Griswold, David~R. Hanson, and John~T. Korb.
\newblock {Generators in Icon}.
\newblock {\em ACM Trans. Program. Lang. Syst.}, 3(2):144--161, 1981.

\bibitem{DBLP:conf/iclp/GuptaBMSM07}
Gopal Gupta, Ajay Bansal, Richard Min, Luke Simon, and Ajay Mallya.
\newblock Coinductive logic programming and its applications.
\newblock In Ver{\'o}nica Dahl and Ilkka Niemel{\"a}, editors, {\em ICLP},
  volume 4670 of {\em Lecture Notes in Computer Science}, pages 27--44.
  Springer, 2007.

\bibitem{DBLP:journals/jfp/Hutton99}
Graham Hutton.
\newblock {A Tutorial on the Universality and Expressiveness of Fold}.
\newblock {\em J. Funct. Program.}, 9(4):355--372, 1999.

\bibitem{yieldProlog}
{Jeff Thompson}.
\newblock {Yield Prolog}.
\newblock Project URL http://yieldprolog.sourceforge.net.

\bibitem{DBLP:books/sp/Liskov81}
Barbara Liskov, Russell~R. Atkinson, Toby Bloom, J.~Eliot~B. Moss, Craig
  Schaffert, Robert Scheifler, and Alan Snyder.
\newblock {\em {CLU Reference Manual}}, volume 114 of {\em Lecture Notes in
  Computer Science}.
\newblock Springer, 1981.

\bibitem{DBLP:conf/popl/LiuKM06}
Jed Liu, Aaron Kimball, and Andrew~C. Myers.
\newblock Interruptible iterators.
\newblock In J.~Gregory Morrisett and Simon L.~Peyton Jones, editors, {\em
  POPL}, pages 283--294. ACM, 2006.

\bibitem{DBLP:conf/padl/LiuM03}
Jed Liu and Andrew~C. Myers.
\newblock {JMatch: Iterable Abstract Pattern Matching for Java}.
\newblock In Ver{\'o}nica Dahl and Philip Wadler, editors, {\em PADL}, volume
  2562 of {\em Lecture Notes in Computer Science}, pages 110--127. Springer,
  2003.

\bibitem{conf/fp/MarlowW92}
Simon Marlow and Philip Wadler.
\newblock Deforestation for higher-order functions.
\newblock In John Launchbury and Patrick~M. Sansom, editors, {\em Functional
  Programming}, Workshops in Computing, pages 154--165. Springer, 1992.

\bibitem{matz}
Yukihiro Matsumoto.
\newblock {The Ruby Programming Language}.
\newblock June 2000.

\bibitem{DBLP:conf/atal/MayfieldLF95}
James Mayfield, Yannis Labrou, and Timothy~W. Finin.
\newblock {Evaluation of KQML as an Agent Communication Language}.
\newblock In Michael Wooldridge, J{\"o}rg~P. M{\"u}ller, and Milind Tambe,
  editors, {\em ATAL}, volume 1037 of {\em Lecture Notes in Computer Science},
  pages 347--360. Springer, 1995.

\bibitem{CSharp}
{Microsoft Corp.}
\newblock {Visual \verb~C#~}.
\newblock Project URL http://msdn.microsoft.com/vcsharp.

\bibitem{moggi:monads}
Eugenio Moggi.
\newblock Notions of computation and monads.
\newblock {\em Information and Computation}, 93:55--92, 1991.

\bibitem{haskell:98rr}
Simon~L. Peyton~Jones, editor.
\newblock {\em Haskell 98 Language and Libraries: The Revised Report}.
\newblock September 2002.
\newblock http://haskell.org/definition/haskell98-report.pdf.

\bibitem{DBLP:conf/oopsla/Sasada05}
Koichi Sasada.
\newblock {YARV: yet another RubyVM: innovating the ruby interpreter}.
\newblock In Ralph Johnson and Richard~P. Gabriel, editors, {\em OOPSLA
  Companion}, pages 158--159. ACM, 2005.

\bibitem{DBLP:conf/icalp/SimonBMG07}
Luke Simon, Ajay Bansal, Ajay Mallya, and Gopal Gupta.
\newblock Co-logic programming: Extending logic programming with coinduction.
\newblock In Lars Arge, Christian Cachin, Tomasz Jurdzinski, and Andrzej
  Tarlecki, editors, {\em ICALP}, volume 4596 of {\em Lecture Notes in Computer
  Science}, pages 472--483. Springer, 2007.

\bibitem{Tarau91:JAP}
Paul Tarau.
\newblock A {S}implified {A}bstract {M}achine for the {E}xecution of {B}inary
  {M}etaprograms.
\newblock In {\em Proceedings of the Logic Programming Conference'91}, pages
  119--128. ICOT, Tokyo, 7 1991.

\bibitem{tarau:cl2000}
Paul Tarau.
\newblock {Fluents: A Refactoring of Prolog for Uniform Reflection and
  Interoperation with External Objects}.
\newblock In John Lloyd, editor, {\em {Computational Logic--CL 2000: First
  International Conference}}, London, UK, July 2000.
\newblock LNCS 1861, Springer-Verlag.

\bibitem{ciclops:jinni}
Paul Tarau.
\newblock {Orthogonal Language Constructs for Agent Oriented Logic
  Programming}.
\newblock In Manuel Caro and Jose~F. Morales, editors, {\em {Proceedings of
  CICLOPS 2004, Fourth Colloquium on Implementation of Constraint and Logic
  Programming Systems}}, Saint-Malo, France, September 2004.

\bibitem{bp7advanced}
Paul Tarau.
\newblock {BinProlog 11.x Professional Edition: Advanced BinProlog Programming
  and Extensions Guide}.
\newblock Technical report, BinNet Corp., 2006.

\bibitem{j2k_ug}
Paul Tarau.
\newblock {The Jinni 2006 Prolog Compiler: a High Performance Java and .NET
  based Prolog for Object and Agent Oriented Internet Programming}.
\newblock Technical report, BinNet Corp., 2006.

\bibitem{Tarau93:CONS}
Paul Tarau and M.~Boyer.
\newblock Nonstandard {A}nswers of {E}lementary {L}ogic {P}rograms.
\newblock In J.M. Jacquet, editor, {\em Constructing Logic Programs}, pages
  279--300. J.Wiley, 1993.

\bibitem{td:tlp}
Paul Tarau and Veronica Dahl.
\newblock {High-Level Networking with Mobile Code and First Order
  AND-Continuations}.
\newblock {\em {Theory and Practice of Logic Programming}}, 1(3):359--380, May
  2001.
\newblock Cambridge University Press.

\bibitem{DBLP:conf/tools/Rossum97}
Guido van Rossum.
\newblock {A Tour of the Python Language}.
\newblock In {\em TOOLS (23)}, page 370. IEEE Computer Society, 1997.

\bibitem{journals/tcs/Wadler90}
Philip Wadler.
\newblock Deforestation: Transforming programs to eliminate trees.
\newblock {\em Theor. Comput. Sci.}, 73(2):231--248, 1990.

\bibitem{wadler92:acm}
Philip Wadler.
\newblock The essence of functional programming.
\newblock In {\em ACM Symposium POPL'92}, pages 1--15. ACM Press, 1992.

\bibitem{wadler:comprehending:lfp:90}
P.L. Wadler.
\newblock Comprehending monads.
\newblock In {\em ACM Conf. Lisp and Functional Programming}, pages 61--78,
  Nice, France, 1990. ACM Press.

\bibitem{DBLP:journals/cj/WegnerE04}
Peter Wegner and Eugene Eberbach.
\newblock {New Models of Computation}.
\newblock {\em Comput. J.}, 47(1):4--9, 2004.

\end{thebibliography}

\end{document}